\documentclass[12pt]{article}

\pdfoutput=1 

\nonstopmode
\usepackage{graphicx,xspace,colortbl}

\usepackage{amsmath,amsthm,amsfonts}
\usepackage{fancybox}

\nonstopmode

\newtheorem{theorem}{Theorem}

\newtheorem{proposition}[theorem]{Proposition}
\newtheorem{corollary}[theorem]{Corollary}

\theoremstyle{definition}
\newtheorem{definition}{Definition}

\newtheorem{assumption}{Assumption}

\newtheorem{remark}[definition]{Remark}

\def\CD{{\cal D}}

\def\CF{{\cal F}}
\def\CG{{\cal G}}

\def\call{\mbox{\rm Call}}
\def\put{\mbox{\rm Put}}

\def\argmin{\mbox{argmin}}

\def\qed{\hfill$\sqcap\kern-8.0pt\hbox{$\sqcup$}$\\}
\def\beq{\begin{eqnarray}}
\def\eeq{\end{eqnarray}}
\def\beqq{\begin{eqnarray*}}
\def\eeqq{\end{eqnarray*}}

\def\be{\begin{equation}}
\def\ee{\end{equation}}

\newcommand{{\X}}{{\bf X}}
\newcommand{{\x}}{{\bf x}}
\newcommand{{\Z}}{{\bf Z}}
\newcommand{{\z}}{{\bf z}}
\newcommand{{\Y}}{{\bf Y}}
\newcommand{{\y}}{{\bf y}}
\newcommand{{\F}}{{\bf F}}
\newcommand{{\bbeta}}{{\bf \beta}}
\newcommand{{\bsigma}}{{\bf \sigma}}
\newcommand{{\bL}}{{\bf L}}
\newcommand{{\bW}}{{\bf W}}
\newcommand{{\bu}}{{\bf u}}
\newcommand{{\im}}{\mbox{Im}}

\def\CT{{\cal T}}

\title{Two-factor capital structure models for equity and credit}
\author{T. R. Hurd\thanks{{Research supported by the
Natural Sciences and Engineering Research Council of Canada and
MITACS, Mathematics of Information Technology and Complex Systems
Canada}}\!\ \ \   and Zhuowei Zhou\\ Dept. of Mathematics and Statistics\\
McMaster
University\\Hamilton ON L8S 4K1\\Canada}
\date{\today}

\begin{document}
\maketitle

\begin{abstract} 

We extend the now classic structural credit modeling approach of Black and Cox to a class of ``two-factor'' models that unify equity securities such as options written on the stock price, and credit products like bonds and credit default swaps. In our approach, the two sides of the stylized balance sheet of a firm, namely the asset value and debt value, are assumed to follow a two dimensional Markov process. Amongst  models of this type we find examples that lead to derivative pricing formulas that are capable of reproducing the main features of well known equity models such as the variance gamma model, and at the same time reproducing the stylized facts about default stemming from structural models of credit risk. Moreover, in contrast to one-factor structural models, these models allow for much more flexible dependence between equity and credit markets. Two main technical obstacles to efficient implementation of these pricing formulas are overcome in our paper. The first obstacle stems from the barrier condition implied by the non-default of the firm, and is overcome by the idea of time-changing Brownian motion in a way that preserves the reflection principle for Brownian motion. The second obstacle is the difficulty of computing spread options: this is overcome by using results in recent papers that make efficient use of the two dimensional Fast Fourier Transform. \end{abstract}


\section{Introduction }\label{introduction}

Merton \cite{Merton74}, Black and Cox \cite{BlacCox76}, and other pioneering
researchers in credit risk well understood that dynamics of a firm's equity and
debt should be modeled jointly and that credit derivatives and equity
derivatives are linked inextricably. To this day, however, it has proved
difficult to capture the dynamical essence of these two aspects of a firm's
capitalization.  The papers by Leland \cite{Leland94} and Leland and Toft
\cite{LelaToft96} provide a conceptual basis, but they remain strongly attached
to the framework of diffusion processes and have a one dimensional source of
randomness . 

The above structural models can all be classified as one-factor models with the
asset process as the only underlying source of randomness.  Such models have the
severe limitation that the firm's equity and debt are perfectly correlated (i.e.
they are related by a deterministic function),  while it is clear in real life
that firms, to a greater or lesser extent, have stochastic liabilities that are
not perfectly correlated with assets. As an extreme illustration, hedge funds, 
with their long/short positions, typically have liabilities as volatile as their
assets. Thus the fact is clear that to accurately model capital structure,
a stochastic factor process of dimension at least two is necessary. In the
context of continuous time finance, the technical and
computational challenges implied by this fact have not yet been satisfactorily addressed, and these challenges are the main focus of the present paper.   Only
a few authors have been able to make substantial headway in modeling actual
observed capital structures by two factor models. Eberlein and Madan
\cite{EberMada10}, in a recent working paper, have treated firm asset and
liabilities as imperfectly correlated processes. Equity equals the
asset/liability spread, and they use this fact to calibrate both the asset and
liability values from the observations of a firm's implied equity volatility
surface.

A second deficiency in the standard structural framework is the reliance on
diffusion processes, with the consequence that default events are predictable
and so instantaneous default is either certain or impossible \cite{Scho03}. Thus
in such models the short spreads are either infinity or zero, counter to the
fact that short spreads are observed to be positive even for investment grade
firms. The natural way to overcome this deficiency is to introduce jumps into
the asset process. A number of authors,  notably
\cite{Baxter06b},\cite{CariScho07},\ have successfully implemented jump diffusion
and pure jump versions of the Merton model. However they share that model's
unrealistically simple debt structure. Similar extensions to the Black-Cox first
passage framework, however, have had only a limited success, due to the
technical difficulty of solving the first passage problem for jump processes.
The Kou-Wang model \cite{KouWang03} with exponentially distributed jumps was
able to work because of the special nature of the underlying process.

This difficulty with predictable defaults was the original motivation for
replacing structural models by reduced form models \cite{JarrTurn95} and
incomplete information models \cite{DuffLand01}.  Recently, a class of
``hybrid'' reduced form models that include the stock price and a default hazard process  have been developed. These model equity and debt
products more realistically by allowing the stock price to jump to zero at the
time of default.  Carr and Wu \cite{CarrWu09} take a stochastic volatility model
for the stock price and assume that the default arrival rate is driven by the
volatility and another independent credit risk factor. In Carr and Linetsky
\cite{CarrLine06}, the stock price has a local volatility with constant
elasticity of variance, and the default intensity is specified as an affine
function of the instantaneous variance of the stock. \cite{MendCarrLine10} obtain even more generality while retaining analytic tractability by applying a time change to the hazard rate and stock price processes. All three frameworks are able to
capture the so called leverage effect and the co-movement between volatility and
default intensity.

Within the structural credit modeling framework, \cite{Hurd09} generalizes
the Black-Cox model by treating the log-leverage ratio
$X_t:=\log V_t/K(t)$ as a time-changed Brownian motion (TCBM), where
$e^{-rt}V_t:=e^{v_t}$ denotes the (per-share) firm asset value process
discounted at a constant rate $r$ and $K(t)$ is a deterministic default
threshold. The time of default is the first passage time for the log-leverage
ratio to cross zero. Like other structural approaches along these lines, this model cure
the inconsistency with observed short spreads and add the flexibility to include
jumps to default and volatility clustering.  One contribution of the TCBM approach in \cite{Hurd09} lies
in an innovative mathematical treatment of
the first passage to default that allows the reflection principle and corresponding first passage formulas for Brownian motion to extend to a broader class of processes, leading to analytical tractability in a more general setting.

The object of the present paper is demonstrate how to embed the Black-Cox framework in a simple  way into a two-factor framework that allows the firm's equity and debt to be
partially correlated while retaining tractability of the underlying default model. We do this by treating the default threshold $K$ as a
positive stochastic process $D_t:=e^{rt+d_t}$ that we can think of as the market value of the firm's liabilities, per share. Put another way, we treat the
firm's debt or liability as a new stochastic process, not fully correlated with
the asset process $V_t$.  If we consider as well the stock price $S_t$ (assuming the number of shares is constant and the firm pays no dividends) and the
log-leverage $X_t$, then a minimal set of additional assumptions for combining these processes is: 

\begin{assumption} The pre-default dynamics of any two of the four processes
$V_t, D_t, S_t,X_t$ is Markovian and time-homogeneous, and determines the
dynamics of the remaining two processes by the equations
\be\label{firmeqns} S_t=\max(V_t-D_t,0),\quad X_t=\log V_t/D_t.
\ee
We assume that the discounted processes $e^{-rt}V_t, e^{-rt}D_t, e^{-rt}S_t$ are
martingales under some risk neutral measure $\mathbb{Q}$, and the interest rate
$r$ is constant\footnote{While the arbitrage pricing theory requires only that
$e^{-rt}S_t$ be a martingale, we make a stronger assumption to simplify the
framework.}. The time of default is \be {t^*}=\inf\{t|X_t\le
0\}=\inf\{t|S_t=0\}, \ee and after default $X_t=S_t=0$. At the time of default,
all securities are assumed to be valued in terms of a ``recovery'' random
variable $R$. 
\end{assumption}

In this paper, we will make additional restrictive assumptions on the form of $V,D,S,X$ to obtain a workable tractable framework. Under these restrictions, the
pure credit dynamics of two factor models with a constant recovery 
will be consistent with the TCBM credit framework of \cite{Hurd09}. In \cite{HurdZhou11a} good
calibrations of such credit models to a large dataset of CDS spreads for Ford Motor Company were obtained, thus verifying the quality of the framework as a model for credit risk. In addition our
two-factor models price equity options as barrier spread options on $V,D$. Thus
option pricing in two factor models faces the type of  computational challenges
for spread options that have been studied in such papers as \cite{DempHong00},
\cite{CarmDurr03} and \cite{HurdZhou11a}. We are able to use the TCBM structure
and properties of Brownian motion to develop an efficient equity option pricing algorithm that
uses a two dimensional Fast Fourier Transform (FFT). Such a fast algorithm is
needed not just in the forward direction for security pricing, but more
importantly  to solve the "inverse problem'' that arises in calibration to a
dataset of observed security prices. In this paper, because we have efficient
pricing of the basic securities to be used in the calibration, we are able to
demonstrate the feasibility of efficient statistical estimation of two factor
models to a dataset of CDS curves and implied equity volatility surfaces. 

The above assumptions on the firm's capital structure can only be valid for a
period $[0,T]$ over which the firm pays no dividends or debt coupons, and does
not issue new shares or debt.  A consistent firm model that incorporates such
real world features will also be of interest for future research. 

This paper will discuss several implementations of the two-factor framework where $S, D$ and $V$
are discounted martingales. In all these implementations, we are able to overcome the two technical obstacles, namely the treatment of the first passage to default and the efficient computation of spread options. In \S 2 and \S 3 we investigate the case where $V_t$
and $D_t$ are correlated geometric Brownian motions (GBMs). The resultant default model
extends the Black-Cox model and shares its known shortcomings, such as zero
short spreads. As well it tends to generate rather unrealistic implied
volatility surfaces. Therefore in \S4 we allow  $V_t$ and $D_t$ to be L\'evy subordinated Brownian motions (LSBMs)
driven by a single time change, in this case either a gamma process or a process with exponential distributed jumps. We investigate
some of the possible shapes of both the CDS curve and the implied vol surface.
In \S5, we investigate how to calibrate such models to market CDS and implied
vol data on a single date. We then exhibit the results of a simple calibration
of the GBM and LSBM models to data for a typical default risky firm,
Ford Motor Company. Finally,  \S6 offers a summary and some directions for
future exploration.

\section{Risk-Neutral Security Valuation}
\label{debtmodel}

As we have explained, the firm's capitalization is modeled by the four processes
$V_t, D_t, X_t, S_t$ satisfying Assumption \eqref{firmeqns}, and default $t^*$ is the first
time $V_t\le D_t$, or equivalently when $S_t=0$. We work in a risk-neutral
filtered probability space $(\Omega,\CF,(\CF_t)_{t\ge 0},\mathbb{Q})$, where $V_t, D_t,
X_t, S_t$ are adapted to the filtration $\CF_t$ and $t^*$ is an $\CF_t$ stopping
time. If $\CG_t\subset\CF_t$ denotes the ``market filtration'', note that the
stock price $S_t$ is $\CG_t$ measurable, whereas $X, V$ and $D$ are not. In
practice $X_t,V_t, D_t$ must be inferred from the prices of securities trading
on the firm, and possibly its quarterly balance sheets.  

In this section, we demonstrate that the price of basic credit and equity
derivatives  can be reduced to computations involving the joint characteristic function
of $Y_t:=[v_T,d_T]$ conditioned on non-default at time $T$:
\be \Phi_{ND}(u_1,u_2;T;v_0,d_0):= \mathbb{E}^\mathbb{Q}_{v_0,d_0}[e^{i(u_1 v_T+u_2
d_T)}{\bf 1}_{\{t^*>T\}}]\ .
\label{PhiNDdef}\ee
As a special case, note that the probability of no default at time $T$ is:
\be\label{nodefault}
P(T;v_0,d_0)=\Phi_{ND}(0,0;T;v_0,d_0)=\mathbb{E}^\mathbb{Q}_{v_0,d_0}[{\bf
1}_{\{t^*>T\}}]\ .
\ee

The reader can anticipate that in subsequent sections, we will introduce a number of  models of dynamics where a generalized reflection principle holds and implies that computations involving
$\Phi_{ND}$ can be reduced to computations involving the unconstrained characteristic function 
\be \Phi(u_1,u_2;T;v_0,d_0):= \mathbb{E}^\mathbb{Q}_{v_0,d_0}[e^{i(u_1 v_T+u_2
d_T)} ]\ .
\label{Phidef}\ee
Moreover,  our models will have a common property on $Y_t$:

\begin{assumption} For any $t>0$, the increment $Y_t-Y_0$ is independent of
$Y_0$.
 \end{assumption}
 \bigskip
 \noindent This implies that the characteristic function $\Phi$ of $Y_T$ factorizes\footnote{Here and subsequently we adopt matrix notation $u=[u_1,u_2]$, $Y_t=[v_t,d_t]$, and in particular $Y'$ denotes the transpose of $Y$.}:
 \begin{equation}\label{charfn}
  {\mathbb{E}}_{T_0}[e^{iuY_T'}]=e^{iuY'_0}\Phi(u;T),  \quad \Phi(u;T):=
{\mathbb{E}}_{Y_0}[e^{iu(Y_T-Y_0)'}]\ .
 \end{equation}
 where $\Phi(u;T)$ is independent of $Y_0$. Thus, in the more specialized setting we have in mind, pricing of all important
derivatives will be reduced to computation of explicit low dimensional integrals.

\subsection{Defaultable Bonds and Credit Default Swaps}
At any time $t$ prior to default ${t^*}$, we consider the value of a zero coupon
bond that pays \$1 at maturity $T$ if the firm is solvent. In the event that ${t^*}<T$, it might be reasonable to suppose
that the recovery value of the bond will be dependent on  $D_{t^*}$. However, to avoid a
detailed analysis of stochastic recovery modeling, we make a mathematically
simple hypothesis:
\begin{assumption} The recovery value of a zero coupon bond with maturity $T$,
at the time of default  ${t^*}<T$ is a constant $R\in[0,1)$.
\end{assumption}

This assumption is analogous to the recovery of par mechanism often made in
credit risk modeling, and it shares some of its limitations. Then one has the
risk-neutral valuation formula for the pre-default  price $B_t(T)$ at time $t$
of the maturity $T$ zero coupon bond:
\be\label{bondformula} 
B_t(T)= \mathbb{E}^\mathbb{Q}_{v_0,d_0}\left[e^{-r(T-t)}{\bf
1}_{\{{t^*}>T\}}+R{\bf 1}_{\{t<{t^*}\le
T\}}e^{-r({t^*}-t)}|\CF_t  \right]\ ,\ee
which leads to:
\begin{proposition} 
\begin{enumerate}
\item The pre-default price at time $t$ of a zero coupon bond with
maturity $T$ with recovery of par is given by
\be
B_t(T)=e^{-r(T-t)}P(T-t;v_t,d_t) + R e^{-r(T-t)}(1-P(T-t;v_t,d_t))\ .
\ee
\item The fair swap rate\footnote{CDS contracts were restructured at the ``big bang'' on 
April 8, 2009, and since then quoted CDS spreads must be reinterpreted. We ignore this complication here, but see \cite{BrigoCDSnote09} for a discussion.}  for a CDS contract with maturity $T=N\Delta t$, with premiums paid in arrears on dates $t_k=k\Delta t, k=1,\dots, N$, and the default payment of $(1-R)$ paid at the end of the period when default occurs, is given by
\be
CDS(T;v_0,d_0)=\frac{(1-R)\left[ \sum_{k=1}^{N-1}[1-P(t_k;v_0,d_0)][e^{-rt_k}-e^{-rt_{k+1}}]+e^{-rT}[1-P(T;v_0,d_0)]\right]}{\Delta t\sum_{k=1}^NP(t_k;v_0,d_0)e^{-rt_k}}\ .\label{CDSformula}
\ee

\end{enumerate}
\end{proposition}

\subsection{Equity derivatives}
We have assumed that $S_t=0$ for all $t\ge {t^*}$. This is a plausible
idealization of the observed fact that stocks typically trade near zero for a
period after a default plus the fact of limited liability that ensures $S_t\ge
0$. By the martingale assumption \label{mgass} it follows that for any $t\le s$
prior to default 
\be S_t= \mathbb{E}^\mathbb{Q}[e^{-r(s-t)}(V_s-D_s){\bf 1}_{\{{t^*}>s\}}|\CF_t ]
=(V_t-D_t){\bf 1}_{\{{t^*}>t\}}.
\label{eqform}
\ee
The second equality comes from Doob's optional stopping theorem. We
notice that 
\[ e^{-rs}(V_s-D_s){\bf 1}_{\{{t^*}>s\}}=e^{-r(s\wedge {t^*})}(V_{s\wedge {t^*}}-D_{s\wedge {t^*}})
\]
 is a $\mathbb{Q}$ martingale
evaluated at a bounded stopping time $s\wedge t^*$, which is also a $\mathbb{Q}$ martingale. 
In \eqref{eqform}, $S_t$ is independent of the debt maturity $s$. This is different from the
standard setup in the Merton model and Black-Cox model, which makes it more parsimonious. 
Moreover, the time $t$ price of a
maturity $T>t$ forward contract with strike $K$ will be $S_t-e^{-r(T-t)}K$. 
A European call option with (positive) strike $K$ and maturity $T$ has time $t$
pre-default value
\be\label{call} \call^{KT}_t= \mathbb{E}^\mathbb{Q}[e^{-r(T-t)}(V_T-D_T-
K)^+{\bf 1}_{\{{t^*}>T\}}|\CF_t].
\ee
Observe that this is equivalent to a down-and-out barrier spread option
with a leverage barrier on $X_t=0$. Put-call parity also holds in such a model,
implying that $ \call^{KT}_t-\put^{KT}_t=S_t-Ke^{-r(T-t)}.$

When a closed or computable form exists for the non-default characteristic function $\Phi_{ND}$ the above option pricing formula is amenable to Fourier analysis, following the
method developed in \cite{HurdZhou10} for vanilla spread options. There it is
proved that the spread option payoff function has an explicit two-dimensional
Fourier transform:
\begin{proposition} \label{mainprop }
 For any real numbers $\epsilon=(\epsilon_1,\epsilon_2)$ with $\epsilon_2>0$ and
$\epsilon_1+\epsilon_2<-1$
 \be (e^{x_1}-e^{x_2}-1)^+=(2\pi)^{-2}\iint_{\mathbb{R}^2+i\epsilon}
e^{i(u_1x_1+u_2x_2)} \hat P(u_1,u_2) d^2u
\label{pfft}
 \ee
 where $\hat
P(u_1,u_2)=\frac{\Gamma(i(u_1+u_2)-1)\Gamma(-iu_2)}{\Gamma(iu_1+1)}$, where
$\Gamma(z)$ is the complex gamma function defined for $\Re z>0$ by the integral 
 \[\Gamma(z)=\int^\infty_0e^{-t}t^{z-1}dt\ .
 \]
 \end{proposition}
Combining this formula with the Fubini Theorem leads to the following formula for a call option with strike $K=1$ and maturity $T$:
 \be
 \call^{T}(v_0,d_0)=\frac{e^{-rT}}{(2\pi)^{2}}\iint_{\mathbb{R}^2+i\epsilon}
\Phi_{ND}(u_1,u_2;T;v_0,d_0)\hat P(u_1,u_2) d^2u\ .
\label{spread}
 \ee
For a general strike $K=e^k$, we use homogeneity to write
 \[\call^{KT}(v_0,d_0)=K\call^{T}(v_0-k,d_0-k).\]
Such explicit double integrals are sometimes efficiently computable for a full range of
$v_0,d_0$ values using a single two-dimensional Fast Fourier Transform.

\section{Geometric Brownian Motion Hybrid Model}
We now consider the two factor model where $V_t=e^{rt+v_t},D_t=e^{rt+d_t}$ are jointly given by a
two-dimensional geometric Brownian motion:
\be \frac{dV_t}{V_t}=rdt+\sigma_v dW_t,\quad \frac{dD_t}{D_t}=rdt +\sigma_d
dZ_t;\quad dW_tdZ_t=\rho dt.\ee
In this case, the stock price $S_t=V_t-D_t$ and log-leverage ratio $X_t=v_t-d_t$ follow the SDEs
\beq
\frac{dS_t}{S_t}&=&\frac{dV_t-dD_t}{V_t-D_t}\\
dX_t&=&-\frac12(\sigma_v^2-\sigma_d^2) dt+\sigma_v dW_t-\sigma_d dZ_t \ .\nonumber
\eeq 
Intuitively one normally expects to find $\sigma_v>\sigma_d\ge 0$, and to have
the correlation $\rho\in(-1,1)$.

\subsection{Stochastic Volatility Model}
Before investigating the form of $\Phi$ and hence the pricing formulas, it is worthwhile to note that this two factor model is identical to a specific stochastic volatility equity model,  analogous to the Heston model. To see this, first we note that we can write 
\be\label{X} X_t=X_0 +\sigma_X[\alpha t+ B_t]\ee
Here $\sigma_X^2=\sigma_v^2-2\rho\sigma_v\sigma_d+\sigma_d^2$ and  $\alpha=\frac{\sigma_d^2-\sigma_v^2}{2\sigma_X}$, and the Brownian motion $B$ is correlated to $W,Z$ with 
\beqq dB dW&=&\rho_{vX}dt,\quad {\sigma_X} \rho_{vX}= \sigma_v-\sigma_d\rho\\
dBdZ&=&\rho_{dX}dt, \quad{\sigma_X} \rho_{dX}=\rho \sigma_v-\sigma_d \ .
\eeqq
 Next we apply the It\^o formula to obtain the SDE for the pre-default stock price
\beq
\frac{dS_t}{S_t}
&=&\frac{r(e^{v_t}-e^{d_t})dt+(\sigma_ve^{v_t}dW_t-\sigma_de^{d_t}dZ_t)}{e^{v_t}
-e^{d_t}}\nonumber\\
&=&rdt+\frac{\sigma_ve^{X_t}dW_t-\sigma_d dZ_t}{e^{X_t}-1}.\nonumber
\label{eqsde}
\eeq 
The martingale term has stochastic quadratic variation with increment $dS^2_t/S^2_t=f(X_t)dt$ where
\be \label{fdef}
f(x):=\frac{(\sigma_ve^x-\sigma_d\rho)^2+(1-\rho^2)\sigma_d^2}{(e^x-1)^2}.
\ee Furthermore, the cross variation increment is $ dX_t dS_t/S_t=g(X_t)\sigma_X\sqrt{f(X_t)} dt$ where \be\label{gdef}
g(x):=\frac{\sigma_v^2e^x-\rho\sigma_v\sigma_d(e^x+1)+\sigma_d^2}{\sigma_X\sqrt{(\sigma_ve^x-\sigma_d\rho)^2+(1-\rho^2)\sigma_d^2}}.
\ee Therefore, using the L\'evy theorem to define a new Brownian motion, one can prove

 \begin{proposition} In the GBM hybrid model, there are independent Brownian motions $B,B^\perp$ such that the log-leverage process is given by 
 \[X_t=X_0 +\sigma_X[\alpha t+B_t]\]
and the stock price follows a stochastic volatility process 
 \be\label{S}
 dS_t/S_t=rdt+\sigma_t [\rho_{SX,t}dB_t+\bar\rho_{SX,t}dB^\perp_t]
 \ee
 with  \be
 \sigma_t^2=f(X_t),\quad 
 \rho_{SX,t}=g(X_t),\quad  \bar\rho_{SX,t}=\sqrt{1-g(X_t)^2}\ . \ee 
 Moreover, the default time $t^*$ is the first passage time for $X_t$ to cross zero, and is predictable.\end{proposition}

\begin{remark} The processes $v_t, d_t$ can  be expressed in terms of the independent drifting BMs $B_t+\alpha t$ and $B_t^\perp+\alpha^\perp t$ where 
$\alpha^\perp=-\frac{\sigma_v}{2\bar\rho_{vX}}=-\frac{\sigma_d}{2\bar\rho_{dX}}$:
 \beq
 v_t&=&v_0+\sigma_v\left[\rho_{vX}(B_t+\alpha t) +\bar\rho_{vX} (B^\perp_t+\alpha^\perp t)\right]\\
 d_t&=&d_0+\sigma_d\left[\rho_{dX}(B_t+\alpha t) +\bar\rho_{dX} (B^\perp_t+\alpha^\perp t)\right]\nonumber
 \label{CofV}\eeq
 where 
 \[ \bar \rho_{vX}=\sqrt{1-\rho^2}\sigma_d/\sigma_X,\quad \bar \rho_{dX}=\sqrt{1-\rho^2}\sigma_v/\sigma_X
\ . \]
 \end{remark}

Finally, we note that in the GBM hybrid model, the explicit characteristic function is
\be\label{PhiGBM}
 \Phi^{GBM}(u;T,Y_0)=\exp\Bigl[iuY_0-\frac{T}{2}u\Sigma u'-i\frac{uT}2(\sigma_v^2,\sigma_d^2)'\Bigr]  \ee
where $\Sigma=[\sigma_v^2, \rho\sigma_v\sigma_d;\rho\sigma_v\sigma_d, \sigma^2_d]$.

\subsection{Pricing}

Basic securities we need to price, namely, defaultable bonds and equity call options, have payoffs that vanish on the set $\{v_T\le d_T\}$ and are subject to a ``down-and-out'' barrier condition. The next proposition shows how the barrier condition can be easily dealt with for such securities.  
First we note that the linear change of variables $[X_{t}, X^\perp_{t}]'=M Y_t,
Y_t:=[v_t,d_t]'$ for the matrix $M=[1,\ -1; 1, \  m]$ with $m=\frac{\rho\sigma_v\sigma_d-\sigma_v^2}{\rho\sigma_v\sigma_d-\sigma_d^2}$ leads to independence of  $X_{t}=v_t-d_t$ and  $X^\perp_t$. This fact allows us to state and prove the following important result:

\begin{proposition} \label{downinprop } Consider an option with maturity $T$ and bounded payoff function $F(v,d)$ that vanishes on the set $\{v<d\}$. Let $f(v_0,d_0;T)$ denote its value at time $0$.
In the geometric Brownian motion model, the down-and-in barrier option
with initial state $v_0>d_0$ and terminal payoff $F$ is equivalent to a vanilla option with the same payoff, but with
linearly transformed initial state $[\tilde v_0,\tilde d_0]$ and an extra factor. Precisely,
 \be\label{vadin}
f_{DI}(v_0,d_0;T)=e^{-2\alpha(v_0-d_0)/\sigma_X}f(\tilde v_0,\tilde d_0;T)
\ee
where $[\tilde v_t,\tilde d_t]'=R[v_t,d_t]', R=M^{-1}[-1,0;0,1] M$.  \end{proposition}

\noindent{Proof:\ } Note that the matrix $R$ is a skewed reflection matrix, which hints that this result is essentially a consequence of the reflection principle for Brownian motion. 
By  intermediate conditioning,
\[
\mathbb{E}_{v_0,d_0}[F(v_T,d_T){\bf 1}_{\{t^*\le T\}}]=\mathbb{E}_{X^\perp_0}[\mathbb{E}_{X_0}[F(M^{-1}[X_T,X_T^\perp]'){\bf 1}_{\{X_T>0\}}{\bf 1}_{\{t^*\le T\}}|X^\perp]] \ .
\]
For fixed $X^\perp$, the reflection principle  the inner expectation can be written as an integral 
\[ 
\int^\infty_0G(x,X_T^\perp)\frac{e^{-2\alpha X_0/\sigma_X}}{\sigma_X\sqrt{T}}\phi\left(\frac{-X_0-x+\alpha\sigma_X T}{\sigma_X\sqrt{T}}\right)\ dx
\]
where $G(x,y):=F(M^{-1}[x,y]')$. Here we have used a standard result for Brownian motion conditioned on crossing a barrier. The vanilla option with the same payoff can be written

\[
f(v_0,d_0)=\mathbb{E}_{v_0,d_0}[F(v_T,d_T)\}]=\mathbb{E}_{X^\perp_0}[\mathbb{E}_{X_0}[F(M^{-1}[X_T,X_T^\perp]'){\bf 1}_{\{X_T>0\}}|X^\perp]]\\
\]
where the inner expectation equals 
\[ \int^\infty_0G(x,X_T^\perp)\frac{1}{\sigma_X\sqrt{T}}\phi\left(\frac{X_0-x+\alpha\sigma_X T}{\sigma_X\sqrt{T}}\right)dx
\]
The desired result follows by a direct comparison of these two formulas for the inner expectations.

\qed

\begin{corollary}\label{GBMpricing} In the geometric Brownian motion hybrid model
\begin{enumerate}
\item The survival probability by time $T$ can be written
\[ P[t^*>T|{v_0,d_0}]=P[v_T>d_T|{v_0,d_0}]-e^{-2\alpha X_0/\sigma_X}P[v_T>d_T|{\tilde v_0,\tilde d_0}]\]
and the price of a zero-recovery defaultable zero-coupon bond is given by
$e^{-rT}P[t^*>T|{v_0,d_0}]$.
  \item The price of an equity call option can be written 
  \[F(v_0,d_0;T)-e^{-2\alpha X_0/\sigma_X}F({\tilde v_0,\tilde d_0};T)\]
  where $F(v_0,d_0;T)=e^{-rT}\mathbb{E}_{v_0,d_0}[(e^{v_T}-e^{d_T}-K)^+{\bf 1}_{\{v_T>d_T\}}].$
The vanilla call option price with maturity $T$ and strike $K=1$ can be computed by the two-dimensional FFT:
\be
 F(v_0,d_0;T)=\frac{e^{-rT}}{(2\pi)^{2}}\iint_{\mathbb{R}^2+i\epsilon}
\Phi^{GBM}(u_1,u_2;T;v_0,d_0)\hat P(u_1,u_2) d^2u
\label{GBMspread}
 \ee\end{enumerate}
\end{corollary}

\section{L\'evy Subordinated Brownian Motion Hybrid Models}
We have seen that the two-factor GBM model implies that the stock process $S$ is a rather specific stochastic volatility process with continuous paths. Moreover the log-leverage process $X$ is an arithmetic Brownian motion with constant drift, and the resultant Black-Cox credit model is well known to be unable to capture the fine effects in observed credit spreads. 

The time-changed Brownian motion (TCBM) credit framework of \cite{Hurd09} introduces a non-decreasing ``time-change'' process $G_t$ independent of $B$ and replaces the Brownian log-leverage process by its time-change $X_t=X_0+\sigma_X[B_{G_t}+\alpha G_t]$ to create a much richer range of dynamics, allowing for purely discontinuous components as well as ``stochastic volatility''. The relevant notion of default by first-passage of the log-leverage process to zero has been well understood in this setting. A non-decreasing L\'evy process $G_t$ is called a subordinator, and a L\'evy subordinated Brownian motion (LSBM) in general includes purely discontinuous components. Any LSBM $W_{G_t}+\alpha G_t$ has the independent increment property and is  Markov  and therefore we can say it is a one-factor process. An important consequence of this one-factor property is that it excludes stochastic volatility effects that by definition involve further factors. 

The  same time-change ideas can be applied to our two-factor GBM hybrid model, and will provide a dramatic increase in flexibility to match effects observed in market data. To retain the simplicity of two-factor models, we focus here on the LSBM case with a single subordinator $G_t$ and the two uncorrelated drifting Brownian motions $B_{G_t}+\alpha G_t, B^\perp_{G_t}+\alpha^\perp G_t$. We assume the natural filtration
$\CF_t$  contains $\sigma\{G_u, B_v, B^\perp_v: 0\le u\le t, 0\le v\le G_t\}$.

The assumptions underlying the LSBM two-factor hybrid model are:
\begin{assumption}
\begin{enumerate}
\item The time-change process  $G_t$ is a L\'evy subordinator with mean  $\mathbb{E}^Q[G_t]=t$.
\item The log discounted firm value $v_t=rt+\log(V_t)$ and log discounted firm liability
$d_t=rt+\log(D_t)$ are both LSBMs, with
the same time change $G_t$, i.e. 
 \beq
 v_t&=&v_0+\sigma_v\left[\rho_{vX}(B_{G_t}+\alpha {G_t}) +\bar\rho_{vX} (B^\perp_{G_t}+\alpha^\perp {G_t})\right]\\
 d_t&=&d_0+\sigma_d\left[\rho_{dX}(B_{G_t}+\alpha {G_t}) +\bar\rho_{dX} (B^\perp_{G_t}+\alpha^\perp {G_t})\right]\nonumber\ .
 \label{TCBMCofV}\eeq
 Here, the parameters are chosen as in section 3.
 \item The log-leverage ratio $X_t:=\log(V_t/D_t)=X_0+\sigma_X[B_{G_t}+\alpha G_t]$ is also a LSBM, and
$S_t=V_t-D_t$. 
  \item The time of default is $t^*$, the first passage time of the second kind for $X$ to cross zero, defined by
    \be\label{firstpassage} t^{*}=\inf\{t|G_t\ge \tau\}\ee
  where $\tau=\inf\{t|B_t+\alpha t\le -X_0/\sigma_X\}$. 
All processes are stopped at $t^*$.  
  \item The interest rate $r$ and recovery fraction $R$ are assumed constant. 
  \end{enumerate}
\end{assumption}

In the model calibration that follows in Section \ref{calsection} we will consider two specific forms for the subordinator $G_t$:\begin{enumerate}
 \item 
The first type of time change is an exponential (EXP) jump process with
constant drift, that is, $G$ has characteristics $(b,0,\nu)$ where $b\in
(0,1)$ and $\nu(z)=ce^{-z/a}/a, a>0$ on $(0,\infty)$, the L\'evy measure, has
support on $\mathbb{R}^+$. The Laplace exponent of $G_t$ is
 \be\label{psiEXP} \psi^{Exp}(u,t):=-\log E[e^{-u G_t}]=t\left[bu +\frac{acu}{1+au}\right]
  \ee
 and by choosing $a=\frac{1-b}{c}$ the average speed of the time change is normalized to 1;
\item The second type of time change is a variance gamma (VG) process
\cite{MadCarCha98}, that is, $G$ is a gamma process with drift having
characteristics $(b,0,\nu)$ where $b\in (0,1)$ and $\nu(z)=ce^{-z/a}/z, a>0$ on
$(0,\infty)$, the L\'evy measure, has support on $\mathbb{R}^+$. The Laplace exponent of $G_t$ is
 \be\label{psiVG}\psi^{VG}(u,t):=-\log E[e^{-u G_t}]=t[bu+c\log(1+au)]
  \ee  
and by choosing $a=\frac{1-b}{c}$ 
the average time change speed is normalized to 1;
 \end{enumerate}

The practical consequence of the precise way the time-change is introduced, and in particular the associated definition of default $t^*$ as a first passage of the second kind, is that all expectations relevant for the pricing of securities can be done efficiently by iterated expectations. 
For example, we have a simple
formula for the characteristic function of $(v_T,d_T)$:
\beqq \Phi^{LSBM}(u_1,u_2;T,v_0,d_0)&=&\mathbb{E}^\mathbb{Q}[\mathbb{E}_{v_0,d_0}^\mathbb{Q}[e^{i(u_1v_T+u_2d_T)}]|G_T]\\
&=&
\mathbb{E}^\mathbb{Q}[\Phi^{GBM}(u_1,u_2;G_T,v_0,d_0)]\ .
\eeqq
Since $\Phi^{GBM}$ given by \eqref{PhiGBM} has the nice feature that
the $T$ dependence takes an exponential affine form which implies that the GBM
pricing formula easily extends to TCBM with a L\'evy subordinator.

\begin{proposition} Consider an option with maturity $T$ and bounded payoff function $F(v,d)$ that pays only if $t^*>T$. Let $f^{GBM}(v_0,d_0;T)$ denote its value at time $0$ under the GBM hybrid model, and $f^{LSBM}(v_0,d_0;T)$ its value under the LSBM model. Then
\[
f^{LSBM}(v_0,d_0;T)=\mathbb{E}^\mathbb{Q}[f^{GBM}(v_0,d_0;G_T)]
\]
\end{proposition}

\noindent{\bf Proof:\ } If we write $v_T=\tilde v_{G_T}, d_T=\tilde d_{G_T}$ and define $\tau$ as in \eqref{firstpassage}, we know that $f^{GBM}(v_0,d_0;T)=\mathbb{E}^\mathbb{Q}_{v_0,d_0}[F(\tilde v_{T},\tilde d_{T}){\bf 1}_{\{\tau> T\}}]$. Then in the LSBM model,
\beq\label{IVformula}
f^{LSBM}(v_0,d_0;T)&=&\mathbb{E}^\mathbb{Q}[\mathbb{E}^\mathbb{Q}_{v_0,d_0}[F(\tilde v_{G_T},\tilde d_{G_T}){\bf 1}_{\{\tau> G_T\}}|G_T]]\\
&=& \mathbb{E}^\mathbb{Q}[f^{GBM}(v_0,d_0;G_T)]\ .\nonumber
\eeq
\qed

As an important example, we can see that combining the above result with Corollary \ref{GBMpricing} leads to the following formula for the equity call option with maturity $T$ and strike $K=1$ in any LSBM model where the time-change $G$ has Laplace exponent $\psi$:
 \be\label{ImpVolformula}
 F^{LSBM}(v_0,d_0;T)-e^{-2\alpha (v_0-d_0)/\sigma_X}F^{LSBM}({\tilde v_0,\tilde d_0};T)\ee
  where the vanilla spread option price is
\be
 F^{LSBM}(v_0,d_0;T)=\frac{e^{-rT}}{(2\pi)^{2}}\iint_{\mathbb{R}^2+i\epsilon}
\exp\Bigl[iuY_0-\psi\Bigl(u\Sigma u'/2-iu(\sigma_v^2,\sigma_d^2)'/2,T\Bigr)\Bigr] \hat P(u_1,u_2) d^2u\ .
\label{LSBMvanillaspread}
 \ee 

\section{Calibration of LSBM models}
\label{calsection}
The aim of this calibration exercise is to demonstrate that the simple two-factor LSBM hybrid framework is capable of fitting simultaneous market CDS and
implied volatility prices on a firm, in this case Ford Motor Company,  at any moment in time. We chose Ford as an example of a large, highly traded, firm, that has been very near to default in recent years. We do not here attempt a large scale survey of how the model performs for a broad range of firms over different periods of time. However, we will see encouraging results from our small study, that suggest that acquiring and analyzing such a dataset may be worth the considerable expense and effort involved. 

\subsection{Data} 
We observed equity and credit market data for Ford Motor Co. obtained from Bloomberg at two moments during the post credit-crunch period: once on July 14, 2010 and once on February 16, 2011. On these dates we noted:
\begin{enumerate}
\item The stock price was \$11.81 and \$16.05 respectively;
\item Midquote implied volatilities $IV_{D,T}$ for  moneyness\\
$\CD:=\{0.4,0.6,0.8,0.9,0.95,0.975,1,1.025,1.05,1.1,1.2,1.3,1.5\}$ and with
times to maturity
  $\CT:=\{37,65,156,191,555\}$ calendar days on July 14, 2010
  $\CT:=\{30,58,93,121,212,338\}$ calendar days on February 16, 2011;
 \item Midquote CDS spreads $CDS_T$ for tenors $
  \tilde{\CT}:=\{1,2,3,4,5,7,10\}$ years;
\item  US treasury yields for maturities\\
  $\bar{\CT}:=\{1m,3m,6m,1y,2y,3y,5y,7y,10y,20y,30y\}$.
\end{enumerate}
From 2009 to 2011, Ford Motor Co. steadily recovered from its near default during the 2007/08 credit crunch and expectations from the financial market correspondingly rose. This improvement in the firm's fortunes is manifested in an observed decrease of both its implied
volatilities and CDS spreads  between the two observation dates.  

\begin{remark} We found that deep in-the-money (ITM) options are not very liquid and deep out-of-the-money (OTM) options are difficult to control numerical errors as their prices are very close to zero. For very short time to maturity options, our FFT formulas \eqref{GBMspread} and \eqref{LSBMvanillaspread} are subject to higher truncation errors as the integrand does not decay fast enough. Therefore in our calibration, we did not use implied volatility data with extreme moneyness $D={0.4, 1.5}$ and with short time to maturity (TTM) $T={30, 37, 58}$ calendar days. 
\end{remark} 

\subsection{Daily Calibration Method}
To test the GBM, EXP and VG models described above, we performed independent daily calibrations to the data observed on the above two dates, making use of the model formulas \eqref{CDSformula} and \eqref{ImpVolformula}. We mention that we adjusted these formulas to incorporate the yield curve $\hat Y_t$ obtained from a boot-strap of the observed US Treasury yields, by the replacement of all discount factors $e^{-rt}$ with corresponding factors $e^{-\hat Y_t t}$. 

It is natural to assume that stock prices are perfectly liquid and hence are perfect observations of the process   $S_t=e^{-rt}[e^{v_t}-e^{d_t}]$. On the other hand, CDS and equity option markets are much less liquid therefore the $N= 7$ observed CDS spreads and $M\sim 50$ observed implied volatilities are not assumed to
match our model prices exactly. Thus at any moment, $S_t$ is exactly observed, while $X_t$ must be filtered from the market data.  

On any given date $t$, under
these assumptions, the risk neutral parameters for both LSBM models to be calibrated are
$\Theta=(\rho,\sigma_v,\sigma_d,b,c,R,X_t)\in\mathbb{R}\times\mathbb{R}^6$. The GBM hybrid model nests inside both LSBM models for any value of $b$ as the limit with $c=\infty$.

Our method is a relative least-squares minimization of the error between the model and market CDS spreads and implied volatilities observed at a single instant in time.   For each $T\in\tilde\CT$, let $\widehat{CDS}_T$ and $CDS_T(\Theta)$ denote market and model CDS spread with maturity $T$. Similarly, for each $T\in\CT$ and $D\in\CD$ let  $\widehat{IV}_{D,T}$ and $IV_{D,T}(\Theta)$ denote the observed and model implied volatilities with the given $D,T$. The objective function to be minimized is a sum of squared relative errors. We introduce a weight factor between the CDS and IV terms to offset a natural overweighting stemming from the large number of equity securities relative to the credit securities. Without this factor, the IV terms would dominate the calibration and wash out the credit effects we aim to capture. Thus we define the objective function to be
\begin{equation*}
\mathbb{J}(\Theta)= \sum_{T\in\tilde{\CT}}
\frac{|\widehat{CDS}_T-{CDS_T}(\Theta)|^2}{{\widehat{CDS}_T}^2}+\frac{1}{C^2}\sum_{T\in\CT,
D\in\CD} \frac{|\widehat{IV}_{D,T}-{IV_{D,T}}(\Theta)|^2}{{\widehat{IV}_{D,T}}^2}
\end{equation*}
where the subjective value $C^2=7$ is chosen to provide a nice balance between the credit and equity datasets. The model
calibration is required to minimize $\mathbb{J}$ over the domain $\Theta\in\mathbb{R}\times\mathbb{R}^6$:
\begin{equation*}
\hat{\Theta}=\argmin_{\Theta}\mathbb{J}(\Theta)
\end{equation*}

\begin{remark}
The above objective function corresponds to the measurement hypothesis that individual spreads and implied volatilities are observed with independent Gaussian relative errors.  Of course, this is likely far from true since in reality, the raw data has been ``cleaned up'' and transformed by Bloomberg in many different ways. Another important point is that the weight  $C^2=7$ in our objective function was chosen by us subjectively to give a nice balance between the credit and equity datasets. We  tried other two choices of weighting schemes, $C^2=1$ and a version weighted by squared bid/ask spreads. Both these schemes turn out to under-weight CDS data resulting in a good fit to equity options while ruining the fit to CDS spreads.
\end{remark}

The calibration was implemented  on a laptop using standard MATLAB. Minimization of $\mathbb{J}$ was implemented using a quasi-Newton optimization algorithm within MATLAB's function ``fmincon'': typically around 200  evaluations of  $\mathbb{J}$ were needed to find an acceptable minimum.
The model CDS and call option prices entering $\mathbb{J}$ are computed by one and two
dimensional Fast Fourier Transforms using MATLAB's functions ``FFT'' and ``FFT2''. Since each FFT2 is computationally intensive, and calibration involves a great number of evaluations, it was effective to use
interpolation from a single grid of FFT-generated prices to compute the range of option prices across moneyness. We used the truncation and
discretization error analysis in \cite{HurdZhou11a} to
optimize these computations. During the calibration, we found that the
down-and-in barrier option terms in \eqref{IVformula} are always much smaller than the vanilla terms. This can be explained because the credit quality of Ford is reasonably high, and therefore the linear transformation $(\tilde v_t,
\tilde d_t)'=R(v_t, d_t)'$  generates a point $\tilde v_t<\tilde d_t$  equivalent to a deep out-of-the-money vanilla spread option.

\subsection{Calibration Results}
The calibration results for the three models and two dates are shown in Table \ref{calibration1}, and Figures 1- 9 show the model and observed CDS and implied volatility curves in each of these cases.  We see that both the EXP and VG models lead to CDS and IV curves that capture the correct levels and quantitative features of the market data. On the other hand, the GBM hybrid model is able to provide a qualitative, not quantitative, fit to the observed market data on the early date, but fails to capture the curves observed on the later date. As a stochastic volatility model, and unlike the Black-Scholes model, it does generate an implied volatility smile, but the details of the fitted shapes are not very good. 

One important point is that both LSBM models (but not the GBM model) on both dates lead to the implied recovery value $R=0$, at the boundary of the parameter domain and counter to common sense. Since reducing $R$ to zero raises the level of CDS curves while leaving IV curves unchanged, this observation suggests a mismatch of risk neutral pricing between equity and credit markets. If the observed CDS spreads are somewhat high 
relative to the level of the implied volatility
surface, the calibration will be forced to choose a zero recovery
rate in order to narrow the gap as much as possible. In our calibrations it appears that this gap is completely closed with $R=0$.  A similar observation has  been
made and discussed in Carr and Wu \cite{CarrWu09} who find a zero recovery rate for 6 out of 8 companies  studied, leaving a significant gap between model prices and market quotes.

Table \ref{firmvaluations} records the balance sheet entries $V,D,S$ implied by each of these calibrations and, for comparison, the summary asset and liability values taken from Ford's quarterly reports on the dates closest to the calibration dates. Clearly there is no obvious direct correspondence between the balance sheet and market values (which appear to be quite sensitive to the choice of model). It will be a nontrivial task for the future to investigate the extent to which the balance sheet numbers can be predicted by the market implied values.

\section{Conclusions}

We have shown that simple capital structure models that treat the firm value and debt value as
correlated stochastic processes can simultaneously capture key empirical features of prices observed in credit and equity
markets. This framework is a significant extension of the standard structural credit risk framework that treats the firm's debt as deterministic or constant over time. We have shown that the necessary pricing computations for vanilla equity options and CDS spreads reduce to explicit one and two dimensional Fourier transforms, and so implementation of such models is comparable in  difficulty to L\'evy process  models now used routinely by banks. These promising results indicate that it will be worthwhile exploring the performance these models on a variety of firms from different countries and sectors. It will also be of interest to explore improvements that may come by using other types of time change. It is in principle quite straightforward to move beyond the present TCBM model with L\'evy subordinators to a far broader modeling framework of time-changed Brownian motions. A range of
implementation and calibration issues will be encountered when considering general TCBMs, 
notably stochastic volatility and jump effects, and addressing such issues will make this a major research and
development undertaking.  

\bibliographystyle{plain}

\newpage

\begin{table}[h] 
\centering\begin{tabular}{||c||c||c|c||}
\hline \hline
&&2010/07/14&2011/02/16\tabularnewline
\hline 
&$\hat\sigma_v$&0.2433&0.2005\tabularnewline\hline
&$\hat\sigma_d$&0.1344&0.1473\tabularnewline\hline
&$\hat\rho$&-0.0699&-0.0143\tabularnewline\hline
&$\hat b$&0.4966&0.6948\tabularnewline\hline 
VG Model&$\hat{c}$&0.0474&0.0240\tabularnewline
\hline
&$\hat R$&0&0\tabularnewline\hline

&$\hat v$&3.1796&3.3393\tabularnewline
\hline
&$\hat d$&2.5036&2.4973\tabularnewline
\hline
&{\rm RMSE}&0.0804&0.0271\tabularnewline\hline\hline
&$\hat\sigma_v$&0.2502&0.2011\tabularnewline\hline
&$\hat\sigma_d$&0.1324&0.1553\tabularnewline\hline
&$\hat\rho$&-0.1687&-0.0383\tabularnewline\hline
&$\hat b$&0.3700&0.7232\tabularnewline\hline 
Exponential Model&$\hat{c}$&0.0519&0.0416\tabularnewline
\hline
&$\hat R$&0&0\tabularnewline\hline

&$\hat v$&3.2786&3.3248\tabularnewline
\hline
&$\hat d$&2.6898&2.4633\tabularnewline
\hline
&{\rm RMSE}&0.0801&0.0265\tabularnewline\hline\hline
&$\hat\sigma_v$&0.0469&0.0612\tabularnewline\hline
&$\hat\sigma_d$&0.0130&0.0095\tabularnewline\hline
&$\hat\rho$&-0.8175&-0.9508\tabularnewline\hline

Black-Cox Model&$\hat R$&0.1900&0.4225\tabularnewline\hline

&$\hat v$&4.5640&4.4767\tabularnewline
\hline
&$\hat d$&4.4327&4.2752\tabularnewline
\hline
&{\rm RMSE}&0.0983&0.1461\tabularnewline\hline\hline
\end{tabular}

\caption{Parameter estimation results and related statistics for the VG, EXP and
Black-Cox 
models. ``RMSE'' denotes the root mean square relative error and is given by $(\mathbb{J}(\hat\Theta))^{1/2}$ with $C^2=1$.}\label{calibration1}
\end{table}

\begin{table}[h] 
\centering\begin{tabular}{||c||c|c|c|c||}
\hline \hline
&FS&VG&EXP&GBM\tabularnewline
\hline 
Asset&119.0/109.5&82.9/107.2&91.6/105.6&331.1/334.2\tabularnewline
\hline 
Debt&78.3/77.7&42.2/46.2&50.8/44.6&290.3/273.2\tabularnewline\hline\hline
\end{tabular}

\caption{Ford's asset and debt values (in \$ Bn) reported in the nearest quarterly financial statements (June 2010 and March 2011) and estimated from models on July 14 2010 and February 16 2011. On those two dates, the number of Ford shares outstanding was approximately 3450 MM shares and 3800 MM shares respectively. From the financial statements, we take the asset value $V$ to be the total current assets plus half of the total long-term assets, and the debt value $D$ to be  the current liabilities.}\label{firmvaluations}
\end{table}

\newpage

\begin{figure}[ht]
\centering\includegraphics[%
  scale=1]{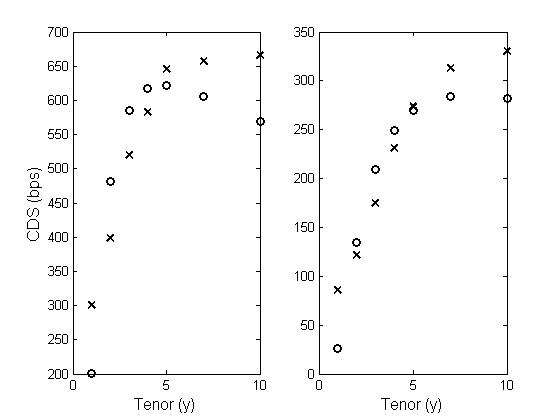}
\caption{{CDS market data (``$\times$'') versus the GBM model data (``$\circ$'') on July 14 2010
(left) and February 16 2011(right).}}
\label{cdsfiggbm}
\end{figure} 

\begin{figure}[ht]
\centering\includegraphics[%
  scale=1.2]{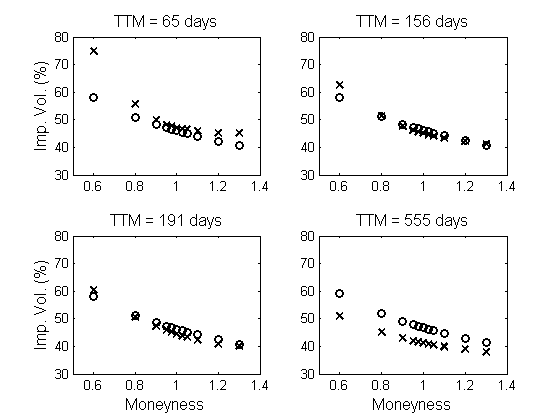}
\caption{{Implied volatility market data (``$\times$'') versus the GBM model data (``$\circ$'') on July 14 2010.}}
\label{GBMimp10}
\end{figure}

\begin{figure}[ht]
\centering\includegraphics[%
  scale=1.2]{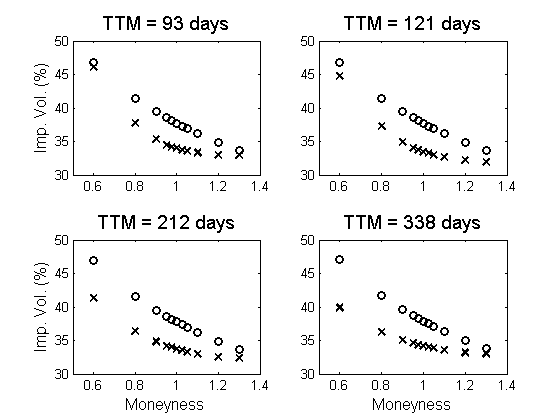}
\caption{{Implied volatility market data (``$\times$'') versus the GBM model data(``$\circ$'') on February 16 2011.}}
\label{GBMimp11}
\end{figure}

\begin{figure}[ht]
\centering\includegraphics[%
  scale=1]{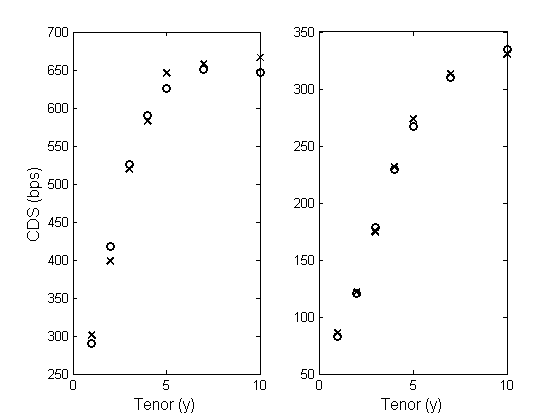}
\caption{{CDS market data (``$\times$'') versus the VG model data (``$\circ$'') on July 14 2010
(left) and February 16 2011(right).}}
\label{cdsfigVG}
\end{figure}

\begin{figure}[ht]
\centering\includegraphics[%
  scale=1.2]{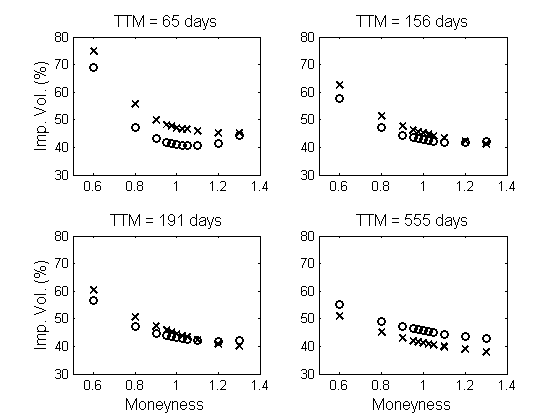}
\caption{{Implied volatility market data (``$\times$'') versus the VG model data (``$\circ$'') on July 14 2010.}}
\label{vgimp1}
\end{figure}

\begin{figure}[ht]
\centering\includegraphics[%
  scale=1.2]{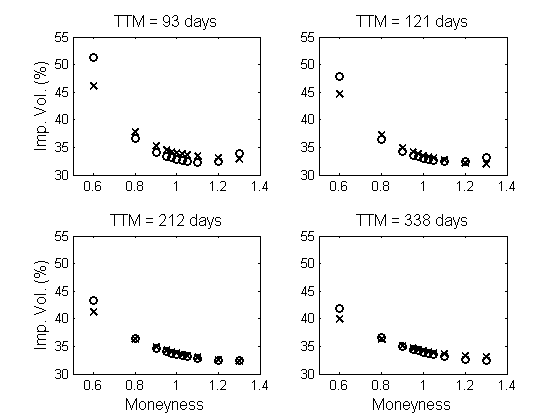}
\caption{{Implied volatility market data (``$\times$'') versus the VG model data (``$\circ$'') on February 16 2011.}}
\label{vgimp2}
\end{figure}

\begin{figure}[ht]
\centering\includegraphics[%
  scale=1]{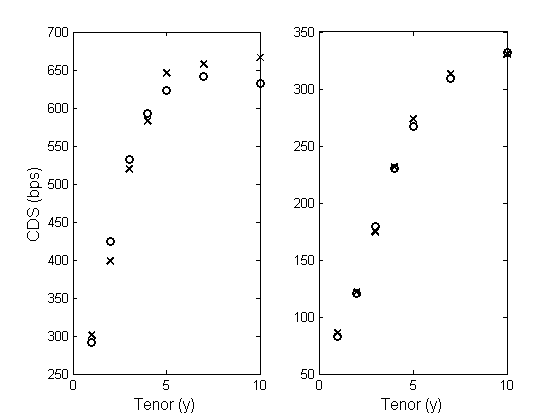}
\caption{{CDS market data (``$\times$'') versus the EXP model data (``$\circ$'') on July 14 2010
(left) and February 16 2011(right).}}
\label{cdsfigEXP}
\end{figure}

\begin{figure}[ht]
\centering\includegraphics[%
  scale=1.2]{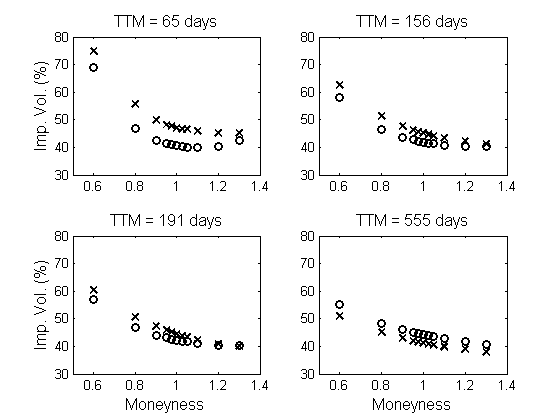}
\caption{{Implied volatility market data (``$\times$'') versus the EXP model data (``$\circ$'') on July 14 2010.}}
\label{imp1}
\end{figure}

\begin{figure}[ht]
\centering\includegraphics[%
  scale=1.2]{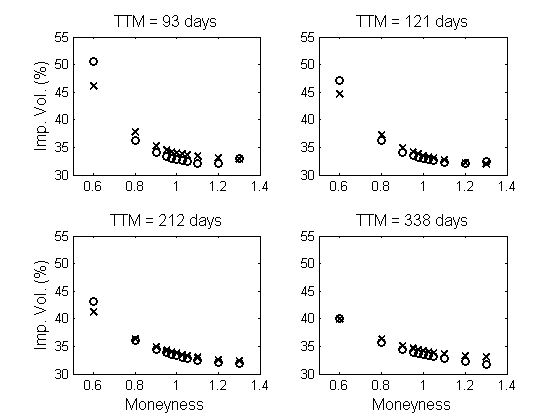}
\caption{{Implied volatility market data (``$\times$'') versus the EXP model data (``$\circ$'') on February 16
2011.}}
\label{imp2}
\end{figure} 

%
%

\end{document}